# On the computation of the entropy for dissipative maps at the edge of chaos using non-extensive statistical mechanics

F. Sattin[1]

*Consorzio RFX, ENEA-Euratom Association, Corso Stati Uniti 4, 35127 Padova, Italy*

**Abstract**

Tsallis' non-extensive statistical mechanics is claimed to be the correct tool to describe the behaviour of low-dimensional dissipative maps at the edge of chaos. Indeed, many different approaches confirm that, for those systems, the evolution is governed by power-laws, not exponential, trends; this coincides with predictions from generalized thermostatistics. In this work, however, we present some analytical considerations, supported also by some simple numerical examples, suggesting the existence of contradictions within this picture.

---

[1] E-mail: sattin@igi.pd.cnr.it

The study of low-dimensional dissipative maps at the edge of chaos has recently attracted much interest. When the control parameter $a$ of the map reaches the value $a = a_c$ at the boundary between ordered and chaotic evolution, Lyapunov exponents are zero and there is still sensitivity to initial conditions, but it is of power-law, not exponential, kind (*weak* sensitivity *versus strong* sensitivity). This is known since some years by numerical calculations [1], while the same conclusions were also reached by different techniques, e.g. Renormalization Group [2]. A review can be found in [3].

Thus, there is little doubt that small-time separation between neighbouring points can be described by a law like

$$\xi_t = \left[1 - (q-1)\lambda_q t\right]^{\frac{1}{q-1}} \qquad . \qquad (1)$$

Such an expression calls obviously for points of contact with Tsallis' non-extensive statistical mechanics [3,4], whose distinctive feature is exactly the prediction of the occurrence of such power-law trends. A study aiming to explicitly establish this connection was recently carried on [5]. For the purposes of the present paper it is useful to briefly summarize its main assumptions and conclusions. The starting point were:

I) Tsallis' re-definition of entropy

$$S_q \equiv \frac{1}{q-1}\left(1 - \sum_i p_i^q\right) \qquad (2)$$

to be opposed to standard Shannon formulation, $S = -\sum_i p_i \ln p_i$ , at which it reduces for $q \to 1$. It is well known that Eq. (2) leads to the non-extensivity for $S_q$, i.e., given two very weakly interacting subsystems $A$, $B$, such that $p_{AB} = p_A p_B$ ,

$$S_q(A+B) = S_q(A) + S_q(B) + (1-q)S_q(A)S_q(B) \qquad . \qquad (3)$$

II) The second essential element used in [5] had been earlier provided by the study [6]. There, it was convincingly demonstrated, for low-dimensional conservative chaotic maps and flows and for intermediate time scales $t$, the connection

$$S \sim k\,t \qquad (4)$$

between the (Shannon) entropy $S$ and the Kolmogorov-Sinai entropy production rate $k$ for a system evolving towards the equilibrium.

In [5] the logistic map was studied

$$x_{t+1} = 1 - a x_t^2 \qquad . \qquad (5)$$

The chaos threshold appears for $a_c = 1.40115519...$ . Numerical simulations showed that in this case Eq. (4) does not hold anymore. Instead, a linear relationship can be recovered provided that $S$ in the l.h.s. of (4) be replaced by its Tsallis' generalization (2) with $q \sim 0.24$. This value is, with good accuracy, the value expected through other ways [2,3]. Therefore, once the validity of II) is assumed also for dissipative systems, the result [5] provides a strong evidence in favour of Tsallis' theory.

Rather unexpectedly, this interpretation seems to leads to some contradictions when this simple picture is only slightly complicated. The purpose of this brief note it to show how and why these problems arise.

Actually, it is straightforward to devise a model where troubles can arise. Let us consider, say, the system made by two weakly coupled logistic maps:

$$x_{t+1} = 1 - ax_t^2 + g(y_t)$$
$$y_{t+1} = 1 - ay_t^2 + g(x_t) \qquad (6)$$

In the following we will refer by $X$ the subsystem which evolves according to the first map of (6), with $Y$ the other subsystem, and with $X+Y$ the whole system. As usual, we can partition the whole $x$ ($y$) range spanned by the maps (6) into discrete intervals and define the corresponding microscopic occupation probabilities $p_x$ ($p_y$). For the whole $X+Y$ system things go exactly the same way, but microstates refer to two dimensional cells. We choose $g(z)$ such that $|g(z)| \ll 1 \ \forall z$, and indeed the limit $g \to 0$ should be borne in mind throughout this work. Therefore, it is reasonable to assume that the probabilities for occupation of the microstates in the two subsystems $X$ and $Y$ are essentially decoupled:

$$p_{X+Y} \approx p_X p_Y \qquad (7)$$

We can compute the entropies for the three systems

$$S_\pi(X+Y) = \frac{1}{\pi-1}\left(1 - \sum_{i,j} p_{ij}^\pi\right) = \frac{1}{\pi-1}\left(1 - \sum_{i,j} p_i^\pi p_j^\pi\right) = k_{XY}t \quad ,$$

$$S_\nu(X) = \frac{1}{\nu-1}\left(1 - \sum_i p_i^\nu\right) = k_X t \quad , \qquad (8)$$

$$S_\nu(Y) = \frac{1}{\nu-1}\left(1 - \sum_j p_j^\nu\right) = k_Y t \quad .$$

In Eq. (8), indices $i$ refer to $X$ subsystem, $j$ to $Y$ subsystem. Suppose to evaluate the middle terms in expressions (8) at the value of the control parameter $a'_c$ ($a'_c \approx a_c$ but they do not need to be necessarily equal because of the interaction term *g*) such that the whole $X+Y$ system is on the edge of chaos. According to point (II) above, it is possible to find a value of the non-extensivity parameter such that the entropies become linear function of time (the right term in each of Eqns. 8). It is not guaranteed that this value must be the same for all the three systems, however, by symmetry considerations, the non-extensivity parameter must be the same for $X$ and $Y$. Therefore, we have used $\pi$ for the whole system and $\nu$ for the two subsystems ($\pi$, $\nu$, can both be different from $q$ as a consequence of the interaction term *g* ).

By exploiting (7) we can expand $S_\pi(X+Y)$:

$$S_\pi(X+Y) = R_\pi(X) + R_\pi(Y) + (1-\pi)R_\pi(X)R_\pi(Y),$$
$$R_\pi(Z) = \frac{1}{\pi-1}\left(1-\sum_z p_z^\pi\right) \tag{9}$$

Formally, the functions $R_\pi$ are equal to the entropies $S_\nu$ but for the replacement $\pi \to \nu$. But it is clear that here we fall into contradiction: let us suppose, infact, $\pi = \nu$. This implies that $R_\pi = S_\nu$. But the r.h.s. of this equality is a linear function of time by definition (Eqns. 8b,c). If we replace it into the first line of (9) we have in the l.h.s. $S_\pi$ which is, still by construction, a linear function of time (Eq. 8a), and in the r.h.s.-instead-appears a quadratic term (because of the term $S_\nu(X)S_\nu(Y)$ ).

There remains the possibility $\pi \neq \nu$, but in this case, according to Tsallis' theory, we should be able to write the entropy of the total system in terms of the entropies $S_\nu$ of the sub-systems from Eqns. (8b,c). Thus, it should be a function of the parameter $\nu$:

$$S_\nu(X+Y) = S_\nu(X) + S_\nu(Y) + (1-\nu)S_\nu(X)S_\nu(Y) \tag{10}$$

The l.h.s. of (10) must be equal to (8a), thus linear in *t* and-again-this is in contradiction with the fact that in the r.h.s. of (10) we have a nonlinear function of time.

We stress that the above considerations are fully self-contained, and should be sufficient to prove the main point of this work. In particular, the validity of Eq. (7) needs a limit $|g| \to 0$, hence is more easily verified within an analytical approach than a numerical one. For completeness, however, we will check these considerations against an actual model: we have performed a numerical study of the system

$$\begin{cases} x_{t+1} = (1-f)(1-ax_t^2) + f y_t \\ y_{t+1} = (1-f)(1-ay_t^2) + f x_t \end{cases} \quad (f=10^{-3}) \tag{11}$$

only slightly different from (6). The calculations were performed along the guidelines of ref. [5]. In order to test our algorithm, we repeated first their computations for the map (5). Since our purpose was just to get a qualitative agreement, we didn't bother of obtaining very accurate estimates; indeed, for the logistic map (5) we did not get an estimate for $q$ better than $0.22 \leq q \leq 0.26$.

The calculations for the system (11) were carried on with an 800×800 mesh, $10^6$ initial conditions, and averaging over 2000 different runs. In order to verify the validity of the hypothesis (7), we computed the quantity

$$\Delta = 2\left\langle \frac{|p_{xy} - p_x p_y|}{p_{xy} + p_x p_y} \right\rangle \qquad (12)$$

where <...> means the average over the runs and the cells, and only the cells where $p_{xy}$ and $p_x p_y$ are not simultaneously zero are included.

Within our accuracy, we found $\pi \approx \nu \approx q$. The transition to chaos still happens around $a_c$, just like for the unperturbed map. The parameter $\Delta$, as expected, is very small, increasing with time and lying in the range $0.01 \div 0.07$ for all $t$ 's considered. Of course, things would be even better, had we chosen a smaller value for $f$.

In conclusion, the interpretation of power-law behaviour of maps at the edge of chaos in terms of non-extensive statistical mechanics can lead to some contradictions. This is consistent with recent works on high-dimensional systems, where it is stated that non-extensive entropy should not be regarded as a fundamental concept.

The author wishes to thank A. Rapisarda for useful comments and suggestions.